# Intrinsic Magnetism of Grain Boundaries in Two-dimensional Metal Dichalcogenides

Zhuhua Zhang,[1] Xiaolong Zou,[1] Vincent H. Crespi,[2] Boris I. Yakobson[1]*

[1]Department of Mechanical Engineering and Materials Science, Department of Chemistry, and the Smalley Institute for Nanoscale Science and Technology, Rice University, Houston, TX 77005, USA.

[2]Department of Physics and Materials Research Institute, The Pennsylvania State University, University Park, Pennsylvania 16802-6300, USA.

*E-mail: biy@rice.edu

Grain boundaries (GBs) are structural imperfections that typically degrade the performance of materials. Here we show that dislocations and GBs in two-dimensional (2D) metal dichalcogenides $MX_2$ (M = Mo, W; X = S, Se) can actually *improve* the material by giving it a qualitatively new physical property: magnetism. The dislocations studied all have a substantial magnetic moment of ~1 Bohr magneton. In contrast, dislocations in other well-studied 2D materials are typically non-magnetic. GBs composed of pentagon-heptagon pairs interact ferromagnetically and transition from semiconductor to half-metal or metal as a function of tilt angle and/or doping level. When the tilt angle exceeds 47° the structural energetics favor square-octagon pairs and the GB becomes an antiferromagnetic semiconductor. These exceptional magnetic properties arise from an interplay of dislocation-induced localized states, doping, and locally unbalanced stoichiometry. Purposeful engineering of topological GBs may be able to convert $MX_2$ into a promising 2D magnetic semiconductor.

Nanostructured magnetic semiconductors have attracted extensive attention [1-2], since they could enable the simultaneous control of spin and charge, promising enhanced efficiency for electronic devices and enriching them with new functions. Following the advent of graphene, there has been an upsurge of interest in the intrinsic magnetism in two-dimensional (2D) crystals. While experiments have detected ferromagnetic signals in certain graphene samples [3-7], these systems show some significant limitations [8]. First, the vanishing band gap of graphene is undesirable for switching functionalities and controllable magnetism. Second, magnetism in graphene appears to originate from vacancies [9-12], adatoms [13-18] or zigzag edges [19,20] — these imperfections can be readily removed by local structural rearrangements through annealing or chemical passivation. Global topological defects such as dislocations within grain boundaries (GBs) are more robust, since they cannot be annealed away by a purely local structural rearrangement. GBs are ubiquitous in 2D materials, forming at the interstices of distinct grains that nucleate during growth [21-24]. In case of graphene, GBs are strings of pentagon-heptagon (5|7) dislocations [24-26]; these are known to be nonmagnetic, as are GBs in h-BN [27,28].

As inorganic analogues of graphene, single-layer transition metal dichalcogenides $MX_2$ (M=Mo,W; X=S,Se) have an intrinsic band gap of 1.4 to 2.0 eV and are superior to graphene in certain respects [29-35]. Unlike graphene, the 2D $MX_2$ family has metal-ligand bonding and a three-atom thickness, with the M atoms sandwiched between layers of X ligands in a trigonal prismatic geometry. The crystal field splitting of M orbitals combined with reduced dimensionality in single-layered $MX_2$ yields a direct band gap [33,34]. Like all crystalline materials, these $MX_2$ are prone to polycrystallinity [36], and recent experiments reveal a variety of dislocations and GBs in $MoS_2$ samples [37-40]. Unlike in graphene or h-BN, dislocations in $MX_2$ have a finite thickness, forming dreidel-shaped polyhedra [36,41]. Little from the extensive study of GBs in graphene [21-26] can be generalized to the $MX_2$ family, where qualitatively new phenomena can be anticipated.

Here we show by first-principles calculations that dislocation cores in 2D $MX_2$ can possess substantial magnetism, resulting from partial occupancy of spin-resolved localized electronic states, with significant spin-spin interactions when the dislocations are aligned into a GB. The magnetic energy, spin ordering, and electronic properties of GBs strongly depend on the GB tilt angle, the type of constituent dislocations, and the doping level. Grain boundaries composed of 5|7 rings behave as a ferromagnetic half-metallic nanowire in a range of moderate tilt while grain boundaries composed of 4|8 cores form antiferromagnetic semiconductors. These results uncover an interesting interplay between dislocation morphology, doping and the spin degrees of freedom and open new avenues for exploring spintronics in semiconducting $MX_2$.

Detailed calculation methods and models are provided in supplemental materials [42]. To establish the generality of the results, we implement grain boundaries and dislocations within three distinct structural models: a $MX_2$ nanoribbon containing a single GB connecting two tilted domains; an extended model with periodic boundary conditions containing a pair of complementary GBs, and a GB loop model (embedded in periodic boundary conditions) with three Mo-rich dislocations [42]. These models isolate and control for the effects of nanoribbon edges and GB dipoles; the ubiquity of magnetism across all three gives confidence in the conclusions.

In projection to the basal plane, the $MX_2$ sheet is isomorphic to graphene or h-BN, with the M atom in one sublattice and superimposed X's in the other [Fig. 1(a)]. A dislocation thus can be designed in $MX_2$ by following the same procedure as in graphene [43]: e.g. removing half of an armchair atomic line and reconnecting seamlessly all of the resultant dangling bonds yields a 5|7 core. The strain energy is proportional to the square of the Burges vector $|\mathbf{b}|^2$; this is lowest for the 5|7 core. In contrast to graphene, bi-elemental $MX_2$ has two types of 5|7 dislocations: a M-rich dislocation with an M-M bond and $\mathbf{b}=(1,0)$ which we call ⊥ and an X-rich dislocation we call ⊤ with an X-X bond and $\mathbf{b}=(0,1)$. Energetically unfavorable homoelemental bonds can be avoided by using a 4|8 core (created by removing two parallel zigzag atomic chains). However, this dislocation has a √3-times larger Burgers vector $\mathbf{b}=(1,1)$. The larger strain energy $\sim|\mathbf{b}|^2$ makes an isolated 4|8 dislocation unlikely, although they can become favorable when compactly aligned in a high-tilt GB, as discussed later.

Considering that most experimental work on GBs in $MX_2$ studies $MoS_2$, we first focus on this material and consider the other $MX_2$ compositions later (they show similar properties). Figures 1(b) and 1(c) present the ground-state magnetization densities of isolated Mo-rich ⊥ and S-rich ⊤ dislocations, as extracted from the ribbon model (the periodic model shows a similar spin splitting and level ordering). This system is actually a low-tilt 9° GB, but at this low dislocation density the defects are well separated. The spin polarization is highly localized at the dislocations, embedded within an ambient 2D semiconducting host. In the Mo-rich ⊥ the spin is mainly on the Mo-Mo bond and its four nearest-neighbor Mo atoms. In the S-rich ⊤ the spin sits mostly on the three Mo atoms of the heptagon, particularly on the two atoms adjacent to the S-S pair. Both dislocation types have total magnetic moments of $\mu = 1.0\ \mu_B$. The energy difference between nonmagnetic and magnetic states is substantial: ~36 meV per dislocation. Magnetism intrinsic to dislocations has not been reported previously, to our knowledge, in any 2D material.

Dislocations are natural constituents of grain boundaries and thus should endow magnetism to them. Here, we consider symmetric GB lines bisecting the tilt angle $\theta$ between adjacent grains, as in Figs. 1(a) and S1 in supplemental materials [42]. An asymmetric GB has higher strain energy than a symmetric one; although not discussed here, they can also be magnetic. Figure 2(a) shows the GB's magnetic moment per unit length $M$ as a function of the tilt angle $\theta$, with proportionality evident below $\theta = 32°$. The linear density of dislocations in low-angle GBs is $2\sin(\theta/2)/b \approx \theta/b$ [44], thus the linear moment density is proportional to the tilt angle: $M \approx \mu\theta/b$, where $\mu$ is the moment of each dislocation. Interestingly, the Mo-rich GBs sustain this proportionality up to a larger tilt angle than do the S-rich GBs: the local moments of the S-rich ⊤ decrease as the S-S bonding geometry becomes more distorted at higher $\theta$; (this distortion may also help explain why S-rich GBs have not yet been found experimentally).

Above 32° the number of 5|7 dislocations declines and hexagons with homoelemental bonds appear instead (symmetric "s-hexagons"). As shown in Fig 2(a), GBs with s-hexagons support further increase in the magnetic moment up to about $\theta = 47°$, but at higher angles the magnetism collapses as the GB approaches a state of pure s-hexagons at $\theta = 60°$. At $\theta = 47°$,

we obtain $M = 1.75$ $\mu_B$/nm for the Mo-rich GB and $M = 1.55$ $\mu_B$/nm for the S-rich GB. These values are significantly higher than the spin density of 1.2 $\mu_B$/nm calculated for zigzag graphene edges. The precise magnetization densities attained in experimental systems will be sensitive to the doping level, as discussed below.

As depicted in Fig. 2(b), above $\theta = 47°$ the 4|8 core becomes more favorable than the 5|7 core: although it has a larger Burgers vector, it has no homoelemental bonds. The prevalence of 8-fold rings at high tilt angles is consistent with recent experimental data on grain boundary structures [37-39]. At $\theta = 60°$ the cancellation of stress fields between adjacent 4|8s lowers the GB energy by ~3.2 eV/nm from that of an array of homoelemental bonds (i.e. all s-hexagons). In contrast, for graphene and h-BN the 5|7s plus hexagons are always preferred over 4|8s. As shown in Fig. 2(a), the magnetic moment of these 4|8 GBs sharply increases as one approaches 60°, reaching 2.1 $\mu_B$/nm in magnitude, which is higher than the maximum achieved by 5|7 boundaries; the spin-polarization energy reaches 45 meV per dislocation. (We quote the magnitudes of local moments which are similar for ferro- and antiferromagnetic states). 4|8s can also appear in antisymmetric zigzag GBs (invariant to mirror reflection with element-inversion). A recently observed GB composed of 8|4|4|8 units [37] which is slanted by 11° from the antisymmetric 60° zigzag GB reaches $M = 1.2 \mu_B$/nm [42]. At high tilt, the strongly magnetic 4|8 and 8|4|4|8 structures surpass the 5|7 dislocation.

What is the strength of the spin coupling? We compare the energies of GBs with ferromagnetic (FM) or anti-ferromagnetic (AFM) order between adjacent dislocations. In GBs composed of 5|7s, the FM structure is energetically preferred. For the Mo-rich grain boundary, $E_{AFM} - E_{FM}$ is ~8 meV for $\theta = 22°$ and increases to ~14 meV for the more closely-spaced dislocations at $\theta = 32°$. In contrast, grain boundaries with 4|8 cores favor an AFM structure: at $\theta = 60°$ we obtain $E_{AFM} - E_{FM} \approx -32$ meV. This distinction may be related to carrier-mediated ferromagnetism [45]: the 5|7 defects support free charge carriers at these doping levels, whereas the 4|8 defects do not, as shown in Fig. 4. Another important distinction between the 5|7 and 4|8 cases are homoelemental bonds in the former but not the latter. Mapping these DFT-computed energies onto a 1D nearest-neighbor Ising model $H = -J\sum_i \mathbf{S}_i \mathbf{S}_{i+1}$ where $J$ is the spin-spin interaction and $\mathbf{S}_i$ is the spin of the $i^{th}$ dislocation (with $S = 1/2$), we obtain $J$ as a function of $\theta$ in Fig. 2(b). For GBs containing 5|7 dislocations, $J$ is positive and increases as the separation between dislocations decreases at larger tilt angle. GBs composed of 4|8s have negative $J$, again larger in magnitude for higher tilt angles. For the most strongly-interacting local moments, at $\theta = 32°$ for FM ordering and $\theta = 60°$ for AFM ordering, the characteristic energy scale for spin fluctuations on the order of room temperature in the AFM case and a significant fraction thereof for the FM case. However, long-ranged magnetic order is not anticipated unless further spin couplings can be introduced to increase the effective dimensionality.

To understand better why dislocations and grain boundaries in $MoS_2$ are magnetic, we examine their electronic structure. Figure 3(a) shows the spin-polarized density of states of the 9° GB described earlier, projected onto just the atoms of the Mo-rich dislocation, compared to that of an ideal defect-free $MoS_2$, shown in gray. Two new localized states – predominately Mo $4d_{x2-y2,xy}$ and $4d_{z2}$ with a small admixture of S $3p_x$ – appear within the bulk band gap, marked by $\delta$ and $\delta^*$. The $\delta$ state has bonding character around the Mo-Mo bond while the $\delta^*$ has antibonding character, as shown in Fig. 3(b). The local excess of Mo weakens the crystal potential slightly, elevating these states slightly in energy relative to the dislocations in the S-rich GB [42]. At an appropriate doping level, these localized states become partially occupied; within a simple Stoner model, when the $\delta$ state crosses the Fermi level, the density of states could trigger spontaneous spin splitting. The spin polarization of the Mo-rich ⊥ thus echoes the spatial distribution of the $\delta$ state, comparing Figs. 1(b) and 3(b). If two S atoms are inserted into the Mo-Mo bond of the Mo-rich ⊥ to balance the local excess of Mo, the resultant 6|8 dislocation is nonmagnetic.

Since grain boundaries comprise a small fraction of a typical experimental 2D system, these dislocation-induced mid-gap states will have a low overall areal density. Thus, the precise location of the Fermi level relative to these mid-gap states will be a sensitive function of the doping conditions. In the calculations, the system is self-doped by the unterminated ribbon edges in the nanoribbon model or by band overlap between complementary dislocations of opposite character in the periodic model. The GB loop model, in contrast, has three identical dislocations and no edge effects: the appearance of magnetism in this system demonstrates that self-doping through band overlap is not required for magnetism. Experiment will provide more fine-grained control of the doping level, through external gating or intrinsic chemical doping. Thus the calculations presented here should be looked upon as a detailed explication of the magnetic response for one particular choice of Fermi energy; calculations [42] of the magnetic properties as a function of doping level (with jellium background) reveal the anticipated tunable magnetic response.

A similar analysis applies for the S-rich ⊤, except with a slight downward shift to the defect states and a spin-polarized antibonding $\delta^*$ state, as shown in Figs. 3(c) and (d). Alleviating the S-excess in the ⊤ by removing two S atoms results in nonmagnetic 4|6 dislocation. The GB composed of 4|8s also lacks homoelemental bonds. However, it does support a magnetic state, albeit one with antiferromagnetic alignment of spins on adjacent dislocations and no net overall spin polarization. The spin polarization again mainly resides on Mo atoms, as shown in the right inset of Fig. 2(a). The Mo-Mo distance is reduced within the GB, compared to in the bulk-like region.

Can the magnetic grain boundaries in MoS$_2$ be half-metallic? With increasing tilt angle the overlap between the electronic states on adjacent dislocations is large; this not only increases the exchange coupling but also broadens the states into a dispersive band. If the defect band for just one spin channel crosses the Fermi level, we obtain a half metal. Figures 4(a) and 4(b) present spin-polarized density of states for the Mo-rich GBs of 22° and 32° tilt within the ribbon model. The spin splitting is larger than the bandwidth, so both systems can support a gap in the spin-down channel with a metallic spin-up channel. Charge transport along the GBs is thus fully spin-polarized. Both the ribbon and the periodic models exhibit half-metallic behavior in the range $13° < \theta \leq 32°$ [42]. The two models show slight differences in the Fermi level placement due to diffferent self-doping conditions; in experiment, this doping level will be gate-controllable. Below this tilt range, the GB is a magnetic semiconductor within our models (e.g. the 9° GB of Fig. 3), since the band dispersion along the GB is too small to metallize any spin channel under our self-doping conditions. Above $\theta = 32°$, states associated with s-hexagons become important and the half-metallic nature is not consistently maintained. At highest $\theta = 60°$, GBs composed of 4|8s are antiferromagnetic semiconductors with a band gap of 0.2 eV, as shown in Fig. 4(c). These results promise that purposeful engineering of topological GBs can tailor spin-dependent transport in 2D $MoS_2$.

The semiconducting bulk state of $MoS_2$ provides the appropriate electronic environment, and the transition metal-ligand bonds then provide the local chemistry necessary to induce magnetism at grain boundaries of $MoS_2$. Magnetic GBs should thus be common within this family of transition metal dichalcogenides, including $WS_2$, $MoSe_2$ and $WSe_2$. In contrast, GBs in π-bonded graphene and h-BN or metallic 1T $MX_2$ are nonmagnetic [42].

In summary, a comprehensive analysis, including first-principles calculations focused on the spatial distribution of spin polarization in states localized to the grain boundaries, reveals intrinsic magnetism for dislocations and grain boundaries in metal dichalcogenides. This coupling of spin and dislocations, facilitated by the semiconducting nature of $MX_2$, the 2D metal-ligand bonding and local unbalanced stoichiometry, opens a new avenue towards 2D magnetic semiconductors through grain boundary engineering.

**FIG. 1 (Color online).** Structure of grain boundary in single-layer MoS$_2$. (a) A symmetric GB with a tilt angle $\theta = 9°$. Each repeat cell includes a Mo-rich ⊥ with a homoelemental Mo-Mo bond, tagged as Mo-rich ⊥. Reversing the grain tilt to -9° creates a S-rich ⊤ with two S-S bonds. Isosurface plots (2×10$^{-3}$ e/Å$^3$) show the magnetization densities of (b) Mo-rich ⊥, and (c) S-rich ⊤. Blue and red colors denote positive and negative values of the magnetization density, respectively.

**FIG. 2 (Color online).** Tilt-dependent magnetism of GBs in MoS$_2$. (a) Magnetic moment $M$ per unit length of GB as a function of tilt angle. All values are obtained from GBs with FM order. Insets show magnetization densities (iso-value 2×10$^{-3}$ e/Å$^3$) of a 38° GB composed of Mo-rich ⊥ and a 60° GB composed of 4|8s. (b) Top left axis: energies of GBs as function of $\theta$, with grey circles for GBs composed of 5|7s and blue squares for GBs composed of 4|8s. Bottom right axis: exchange coupling parameter $J$ for several typical GBs composed of Mo-rich ⊥ or 4|8s.

**FIG. 3 (Color online).** Electronic structure analyses of dislocations in MoS$_2$. (a) Spin-polarized density of states (DOS) of a 9° GB composed of Mo-rich ⊥. Solid lines are the DOS projected onto those atoms forming the GB while grey shaded regions are the DOS projected onto atoms in bulk area. (b) Schematic energy levels of ⊥-induced spin-unresolved localized states, $\delta$ and $\delta^*$, and their corresponding isosurface (5×10$^4$ e/Å$^3$) distributions of partial charge densities. A local potential maximum around the ⊥ is illustrated by a potential profile. (c,d) Corresponding cases for a 9° GB composed of S-rich ⊤, showing a local potential minimum. The single arrow denotes a partially occupied energy level.

**FIG. 4 (Color online).** Tilt-dependent ground state electronic properties of GBs in MoS$_2$. Spin-polarized DOS of (a) 22° and (b) 32° GBs composed of Mo-rich ⊥ as well as (c) a 60° GB composed of 4|8s. The color stipulations are the same as in Fig. 3(a).

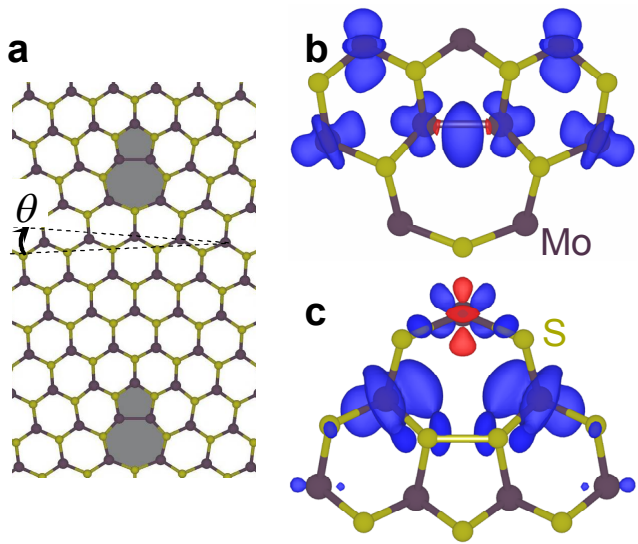

Figure 1

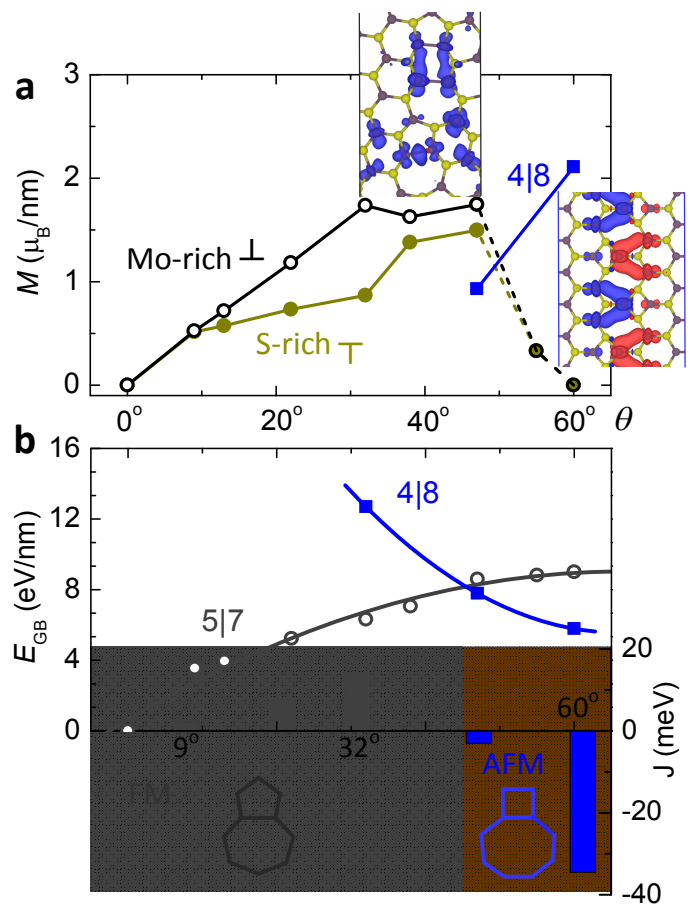

Figure 2

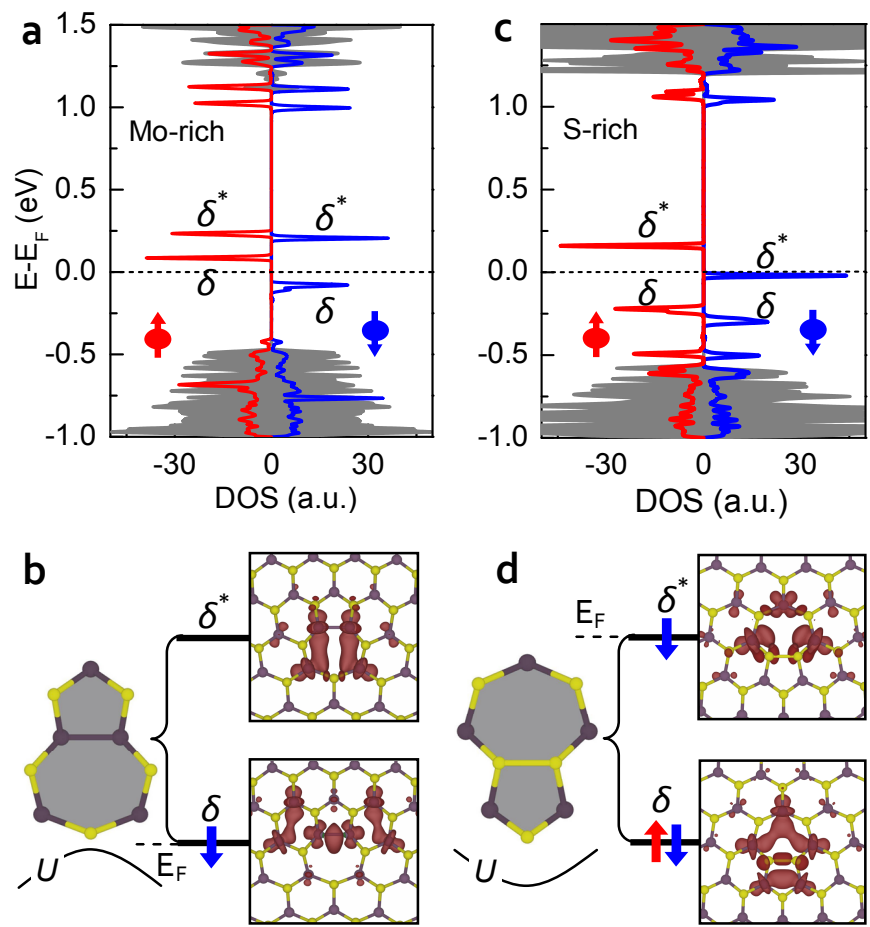

Figure 3

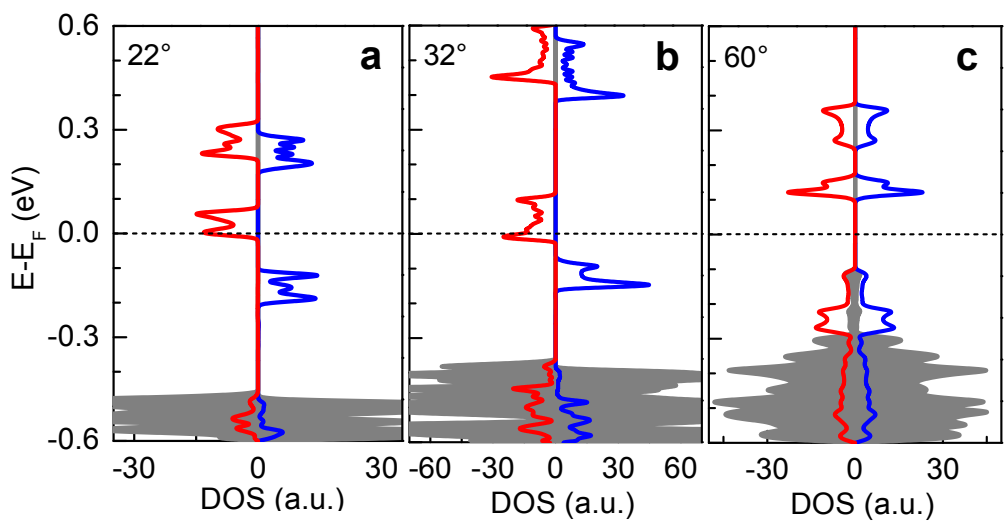

Figure 4